\documentclass{Interspeech}

\interspeechcameraready

\title{Speech-IFEval: Evaluating Instruction-Following and Quantifying Catastrophic Forgetting in Speech-Aware Language Models}

\author[affiliation={1}]{Ke-Han}{Lu}
\author[affiliation={1}]{Chun-Yi}{Kuan}
\author[affiliation={1}]{Hung-yi}{Lee}

\affiliation{Graduate Institute of Communication Engineering}{National Taiwan University}{Taiwan}
\email{d12942024@ntu.edu.tw}

\usepackage{comment}
\usepackage{array, booktabs, multirow, longtable}
\usepackage{amsmath}
\usepackage{cite}
\usepackage{url}
\usepackage{subcaption}
\usepackage{hyperref}
\usepackage{float}
\usepackage{xcolor,colortbl}
\definecolor{Gray}{gray}{0.85}
\newcolumntype{a}{>{\columncolor{Gray}}c}

\begin{document}

\maketitle

\begin{abstract}

We introduce \textbf{Speech-IFEval}, an evaluation framework designed to assess instruction-following capabilities and quantify catastrophic forgetting in speech-aware language models (SLMs). Recent SLMs integrate speech perception with large language models (LLMs), often degrading textual capabilities due to speech-centric training. Existing benchmarks conflate speech perception with instruction-following, hindering evaluation of these distinct skills. To address this gap, we provide a benchmark for diagnosing the instruction-following abilities of SLMs. Our findings show that most SLMs struggle with even basic instructions, performing far worse than text-based LLMs. Additionally, these models are highly sensitive to prompt variations, often yielding inconsistent and unreliable outputs. We highlight core challenges and provide insights to guide future research, emphasizing the need for evaluation beyond task-level metrics.\footnote{https://github.com/kehanlu/Speech-IFEval}

\end{abstract}

\begin{table*}[]
    \caption{Examples of constraints. The constraint prompts are directly appended to the original instruction in closed-ended questions and chain-of-thought (CoT) categories.}
    \centering
    \begin{tabular}{l|l}
        \toprule
        \textbf{Constraint type}  & \textbf{Example} \\
        \midrule
        \textit{Close-ended question} &  \\
        Change case (CC) & Please format your entire response in capital letters only. \\
        Start with / End with (SE) & Start your response with "The answer is:" \\
        Wrap (WR) & Wrap your entire response with double quotation marks. \\
        JSON (JS) & Put your answer in a valid JSON format. \\
        \midrule
        \textit{Creative writing} &  \\
        Bullet points (BP) & List 2 likely reasons why the audio was recorded, formatted as bullet points. \\
        Keywords (KW) & Describe the audio. You must use the word 'spicy'. \\
        Length (LN) & Rewrite the content of the audio as a story with a minimum length of 100 words.\\
        \midrule
        \textit{Chain-of-thought (CoT)} & Analyze this question carefully, step by step. \\
        
        \bottomrule
    \end{tabular}
    
    \label{tab:constraint_prompts}
    \vspace{-1em}
\end{table*}

\section{Introduction}
Recent advances in instruction-following language models (LLMs) have demonstrated remarkable capabilities in processing arbitrary user instructions \cite{achiam2023gpt,bai2023qwen,yang2024qwen25,yang2024qwen2technicalreport,touvron2023llama,dubey2024llama,vicuna2023,team2023gemini}. Speech-aware language models (SLMs) \cite{chu2023qwen,chu2024qwen2, lu2024developing, lu24c_interspeech, tang2024salmonn, gong2023joint, held2024distilling, wang2024blsp, hu2024wavllm,kuan2024speech} extend this success by integrating speech perception into the broad textual knowledge of LLMs. This enables them to process speech input and textual instructions to perform speech-related tasks. 
The development of SLMs revolves around two fundamental capabilities: \textbf{speech perception} and \textbf{instruction-following}. The former involves accurately interpreting speech-related information, such as spoken content, paralinguistic cues, and speaker identities. The latter refers to the model’s ability to comprehend textual directives and execute tasks accordingly, ensuring that responses align with user intent.
To evaluate SLMs, several studies have introduced benchmarks \cite{10448257, yang-etal-2024-air, huang2025dynamicsuperb, chen-etal-2024-beyond-single, gao2024benchmarking, sakshi2025mmau} that pair speech with instructions across various domains. These benchmarks primarily evaluate task-level performance, requiring models to process both speech inputs and textual instructions to generate correct responses.


In the context of SLM development, catastrophic forgetting \cite{goodfellow2013empirical} remains a major challenge\cite{tang2024salmonn, lu24c_interspeech, lu2024developing, hu2024wavllm, yu-etal-2024-self-powered, gong2023joint}. Despite being built on powerful text-based LLMs, SLMs often experience a decline in textual capabilities after training on speech-text pairs. However, existing benchmarks conflate speech perception with instruction-following ability, making independent evaluation of these capabilities difficult. This overlap leads to ambiguous performance assessments.
For example, poor performance may arise from either difficulties in understanding speech or challenges in following textual instructions or understanding questions.


To address this limitation, we propose \textbf{Speech-IFEval}, an evaluation framework focused on assessing the core instruction-following capabilities of SLMs. Unlike existing benchmarks, Speech-IFEval explicitly disentangles instruction-following from speech perception, enabling a more accurate assessment of a model’s ability to interpret and execute textual directives. As part of our benchmark design, we augment speech-instruction pairs with additional output constraints, such as enforcing specific response formats or length requirements (Table~\ref{tab:constraint_prompts}). Importantly, these constraints are independent of the speech input and rely solely on the model’s textual understanding.
Our evaluation focuses on verifying whether models adhere to the prescribed output constraints, thereby isolating instruction-following ability from other confounding factors. Additionally, since SLMs are typically built upon different text-based LLMs, we introduce a cascade framework as a reference system. This setup enables direct comparison between an SLM and its original LLM counterpart, enabling a quantifiable, intra-model analysis of catastrophic forgetting.

Experimental results reveal that most SLMs struggle to follow even basic constraints, whereas text-based reference systems achieve a high success rate with ease. This discrepancy highlights a significant degradation after speech-text training.
Furthermore, when evaluating task-level correctness, we observed notable fluctuations in model accuracy upon the introduction of additional formatting constraints. Some models altered their responses or generated hallucinated content.
These findings suggest that current SLM approaches fail to preserve the intrinsic textual capabilities of LLMs, ultimately compromising their flexibility and robustness in handling arbitrary user input. By revisiting and isolating the underlying factors, our study underscores the importance of evaluating SLMs beyond task-level performance alone.

Our contribution can be summarized as follows:
\begin{itemize}
    \item We introduce Speech-IFEval, an evaluation framework designed to assess the instruction-following capabilities of SLMs and systematically measure the impact of catastrophic forgetting during their development. 
    \item Our findings indicate that most SLMs exhibit significant degradation compared to text-based LLMs and are highly sensitive to variations in instructions, highlighting the limitations of current SLM approaches and evaluation benchmarks.
\end{itemize}



\section{Related works}
LLMs are highly adaptable AI systems capable of interpreting and executing diverse natural language instructions, making them an intuitive interface for human-machine interaction. To assess their instruction-following capabilities, several benchmarks have been developed \cite{zhou2023instruction,qin-etal-2024-infobench, xia-etal-2024-fofo}. One early effort, IFEval \cite{zhou2023instruction}, introduces verifiable constraints within instructions to determine whether models respond as intended. By isolating external knowledge factors, this approach enables the evaluation of fundamental LLM abilities before more comprehensive knowledge-based assessments.


In the context of SLMs, recent studies \cite{10448257, yang-etal-2024-air, huang2025dynamicsuperb, chen-etal-2024-beyond-single, gao2024benchmarking, sakshi2025mmau, chen-etal-2024-beyond-single, wang2024audiobench} have examined performance across speech-related tasks in diverse domains and scenarios. However, these studies primarily emphasize task-level performance, which inherently relies on both instruction-following and speech perception ability, meaning the model must understand the intent and generate an accurate response based on the speech information.
Notably, some benchmarks reveal that certain SLMs struggle to follow expected answer formats \cite{huang2025dynamicsuperb,yang-etal-2024-air, 10448257}, such as selecting a valid option in multiple-choice questions. To address this limitation, researchers have developed LLM-based evaluation \cite{chiang-lee-2023-large} methods to enhance accuracy when interpreting ambiguous responses. Motivated by these observations, Speech-IFEval re-examines fundamental evaluation challenges and provides a comprehensive comparison of recent SLM advancements.

\section{Speech-IFEval}
\begin{table*}[]
    \small	
    \caption{Results of following rate (\%) of SLMs and reference systems. Forgetting rate($\Delta$) is the relative IFrate difference between SLMs and their LLM counterpart.}
    \centering
    \setlength{\tabcolsep}{4pt}
    \begin{tabular}{l|ccccc|cccc|c|cc}
        \toprule
        
        \multirow{2}{*}{Model} & \multicolumn{5}{c|}{\textbf{Closed-ended questions}}  & \multicolumn{4}{c|}{\textbf{Creative writing}} & \textbf{CoT} & \multirow{2}{*}{\textbf{IFrate}} & \multirow{2}{*}{\textbf{$\Delta$}}\\
        & CC & SE & WR & JS & \textbf{ALL} & BP & KW & LN & \textbf{ALL} & \textbf{ALL} & & \\
        \midrule
        \textit{Reference systems} \\
        Vicuna 7B \text{v1.1}  & 22.40 & 71.50 & 29.18 & 88.00 & 52.20 & 93.00 & 73.00 & 73.00 & 78.00 & 64.00 & 64.73 & -- \\
        Vicuna 13B \text{v1.1}  & 74.40 & 68.00 & 60.09 & 85.60 & 72.45 & 91.00 & 74.00 & 74.00 & 78.25 & 71.50 & 74.07 & -- \\
        Qwen-7B-chat & 64.40 & 68.50 & 24.89 & 90.00 & 62.27 & 87.00 & 78.00 & 68.00 & 75.25 & 82.50 & 73.34 & -- \\
        Qwen2-7B-Instruct  & 99.60 & 99.50 & 86.70 & 97.60 & \textbf{95.82} & 97.00 & 79.00 & 84.00 & 86.00 & 67.50 & 83.11 & --  \\
        Qwen2.5-7B-Instruct & 97.60 & 99.50 & 98.28 & 88.40 & 95.71 & 97.00 & 89.00 & 73.50 & 83.25 & 86.50 & 88.49 & -- \\
        Llama2-7B-Chat  & 37.60 & 67.50 & 54.94 & 89.60 & 62.27 & 89.00 & 61.00 & 67.00 & 71.00 & 92.50 & 75.26 & -- \\
        Llama3-8B-Instruct  & 97.20 & 91.50 & 93.56 & 90.80 & 93.35 &	95.00 & 94.00 & 93.00 & \textbf{93.75} & 90.50 & 92.53 & -- \\
        Llama3.1-8B-Instruct  & 92.80 & 89.00 & 76.39 & 94.40 & 88.32 & 97.00 & 93.00 & 92.50 & \textbf{93.75} & \textbf{98.50} & \textbf{93.52} & -- \\
        
        \midrule
        \textit{SLMs} \\
        Qwen-Audio-Chat & 2.40 & 17.50 & 3.43 & 21.20 & 10.93 & 74.00 & 56.00 & 47.00 & 56.00 & 32.00 & 32.98 & -- \\
        Qwen2-Audio-Instruct & 76.80 & 69.50 & 24.46 & 0.00 & 41.59 & 72.00 & 76.00 & 61.50 & 67.75 & 32.00 & 47.11 & -- \\
        
        LTU-AS  & 7.60 & 60.00 & 1.72 & 50.40 & 28.83 & 42.00 & 55.00 & 47.00 & 47.75 & 11.00 & 29.19 & -54.90 \\
        SALMONN  & 54.80 & 38.00 & 18.03 & 37.60 & 37.41 & 64.00 & 63.00 & 59.00 & 61.25 & 12.00 & 36.89 & -50.20 \\
        \quad \quad \textcolor{gray}{($\alpha=16$)}  & \textcolor{gray}{74.40} & \textcolor{gray}{65.00} & \textcolor{gray}{56.22} & \textcolor{gray}{87.20} & \textcolor{gray}{71.28} & \textcolor{gray}{89.00} & \textcolor{gray}{69.00} & \textcolor{gray}{68.00} & \textcolor{gray}{73.50} & \textcolor{gray}{48.00} & \textcolor{gray}{64.26} & \textcolor{gray}{-13.24} \\
        \quad \quad \textcolor{gray}{($\alpha=4$)}  & \textcolor{gray}{67.20} & \textcolor{gray}{68.50} & \textcolor{gray}{54.51} & \textcolor{gray}{84.40} & \textcolor{gray}{68.92} & \textcolor{gray}{92.00} & \textcolor{gray}{65.00} & \textcolor{gray}{76.00} & \textcolor{gray}{77.25} & \textcolor{gray}{58.00} & \textcolor{gray}{68.06} & \textcolor{gray}{-8.12} \\
        \quad \quad \textcolor{gray}{($\alpha=1$)}  & \textcolor{gray}{62.80} & \textcolor{gray}{69.50} & \textcolor{gray}{51.07} & \textcolor{gray}{81.60} & \textcolor{gray}{66.34} & \textcolor{gray}{86.00} & \textcolor{gray}{56.00} & \textcolor{gray}{73.50} & \textcolor{gray}{72.25} & \textcolor{gray}{59.00} & \textcolor{gray}{65.86} & \textcolor{gray}{-11.08} \\
        BLSP-emo  & 61.20 & 59.50 & 45.92 & 96.00 & 66.35 & 74.00 & 59.00 & 61.00 & 63.75 & 50.50 & 60.20 & -17.92 \\
        DiVA & 96.00 & 75.00 & 89.78 & 70.80 & 83.14 & 33.00 & 72.00 & 71.00 & 61.75 & 83.50 & 76.13 & -17.73 \\
        DeSTA2 & 96.40 & 84.50 & 90.13 & 64.40 & \textbf{83.71} & 95.00 & 91.00 & 90.57 & \textbf{92.49} & \textbf{91.50} & \textbf{89.23} & \textbf{-3.57} \\
        
        \bottomrule
    \end{tabular}
    
    \label{tab:main_results}
    \vspace{-1em}
\end{table*}

\subsection{Dataset}


To assess the instruction-following capabilities of SLMs, we propose an evaluation framework that applies output constraints to speech-instruction pairs, such as adhering to specific formats or behaviors. Our benchmarks encompass three core task categories: closed-ended questions, creative writing, and chain-of-thought reasoning. Table \ref{tab:constraint_prompts} presents representative examples.
For a well-designed and user-friendly system, handling such prompts should be seamless and intuitive.



\textbf{Closed-ended questions} Closed-ended questions provide a fundamental assessment of an SLM's ability to handle speech-related tasks. We compile a set of closed-ended questions sourced from established speech-instruction benchmarks \cite{huang2025dynamicsuperb,yang-etal-2024-air,sakshi2025mmau}, covering automatic speech recognition (ASR), speech emotion recognition (SER), and gender recognition (GR). We further integrate general speech understanding tasks from the MMAU speech subset \cite{sakshi2025mmau}.
Each speech-instruction pair is modified by appending a constraint prompt to the original instruction in the format of:
$$
\{\texttt{Instruction}\} \{\texttt{Constraint Prompt}\}
$$
For instance, an ASR instruction may be reformulated as: \textit{"Transcribe the speech. Please format your entire response in capital letters only."} As a result, we assess the model’s ability to follow the constraint rather than the correctness of its transcriptions.

\textbf{Creative writing} We curate speech-related instructions that require models to generate open-ended responses based on a given speech input, such as stories, articles, or descriptive passages, while following the given constraints. This evaluation measures the model's ability to produce well-structured long-form text while maintaining adherence to predefined formatting or stylistic guidelines.

\textbf{Chain-of-thought} Chain-of-thought(CoT) prompting \cite{wei2022chain} is a common technique in LLMs, guiding models to follow a step-by-step reasoning process rather than generating direct answers. The evaluation follows the same construction as the closed-ended questions but with the prompt to encourage sequential reasoning. Instead of expecting a direct response, the model is required to articulate a step-by-step reasoning process. 

We define Instruction-Following Rate (IFrate) as the average following rate across three categories. Higher IFrate indicates a model's stronger capability to follow directives accurately.

\subsection{Reference systems}

Most SLMs are derived from instruction-tuned LLMs to enhance their ability to handle diverse textual instructions. However, since SLMs are built upon different underlying LLMs with varying strengths, direct comparisons of inter-model performance remain challenging.
To enable a more consistent and fair evaluation, we establish a cascade pipeline that leverages a text-based LLM as a reference system, allowing it to process speech inputs in textual form. This pipeline integrates specialized pre-trained speech models for ASR \cite{radford2023robust}\footnote{whisper-large-v3}, SER \cite{ma-etal-2024-emotion2vec}\footnote{emotion2vec-plus-large}, and GR\footnote{\url{https://huggingface.co/alefiury/wav2vec2-large-xlsr-53-gender-recognition-librispeech}}, extracting relevant textual attributes from the speech input. These attributes are then structured into a standardized format, such as: \textit{[00:00:00 - 00:00:05] How are you (Gender: Male, Emotion: Happy)}.
This approach enables direct comparison between the text-based LLM and SLMs under identical instructions. To quantify performance changes, we introduce the forgetting rate ($\Delta$), which measures the relative difference in IFrate, defined as:
$$
\Delta = \frac{\text{IFrate}_\text{SLM} - \text{IFrate}_\text{Ref}}{\text{IFrate}_\text{Ref}}
$$
A high forgetting rate (\(\Delta\)) indicates larger degradation in instruction-following capability after speech-text training.

\section{Experiment setup}
\begin{table}[]
    \small
    \caption{Details of SLMs and their corresponding text-based LLM components. *Qwen-7B is not an instruction-tuned LLM.}
    \centering
    \begin{tabular}{l|l c}
        \toprule
        \textbf{SLM} & \textbf{LLM}\\
        \midrule
        Qwen-Audio-Chat \cite{chu2023qwen} & Qwen-7B* \cite{bai2023qwen} \\
        Qwen2-Audio-Instruct \cite{chu2024qwen2} & Qwen-7B* \cite{bai2023qwen} \\
        LTU-AS \cite{gong2023joint} & Vicuna 7B \text{v1.1} \cite{vicuna2023} \\
        SALMONN \cite{tang2024salmonn} & Vicuna 13B \text{v1.1} \cite{vicuna2023}  \\
        BLSP-emo \cite{wang2024blsp} & Qwen-7B-Chat \cite{bai2023qwen} \\
        DiVA \cite{held2024distilling} & Llama3-8B-Instruct \cite{dubey2024llama} \\
        DeSTA2 \cite{lu2024developing} & Llama3-8B-Instruct \cite{dubey2024llama} \\

        \bottomrule
    \end{tabular}
    \label{tab:model_table}
    \vspace{-1em}
\end{table}

\subsection{Model details}
As shown in Table \ref{tab:model_table}, we evaluate publicly available SLMs \cite{tang2024salmonn, gong2023joint, chu2023qwen, chu2024qwen2, wang2024blsp, lu2024developing, held2024distilling} alongside their corresponding LLM components \cite{vicuna2023, bai2023qwen, dubey2024llama}, which serve as reference systems. We also include different Qwen \cite{yang2024qwen25,yang2024qwen2technicalreport} and Llama\cite{touvron2023llama, dubey2024llama} versions as additional baselines. Specifically, all SLMs share a common architecture, consisting of a speech encoder that extracts speech features and a LLM that jointly processes both speech features and text instructions. All models undergo speech-text training to enable seamless integration of the two modalities.
Unlike other speech language models, the Qwen-Audio series employs Qwen-7B as its backbone LLM, which does not include a text instruction-tuning stage. Instead, it acquires instruction-following capabilities through speech-instruction training. As a result, the Qwen-Audio series lacks a direct reference system in our evaluation.

\begin{table*}[]
    \small
    \caption{Results on task-level performance, under original instructions (left) and instructions with additional constraints (right). The performance of speech models are from the individual speech models used in reference systems.}
    \centering
    \begin{tabular}{l|cc|cc|cc|cc|c}
    \toprule
    \textbf{Model} & \multicolumn{2}{c}{\textbf{ASR $\downarrow$}} & \multicolumn{2}{c}{\textbf{SER $\uparrow$}} & \multicolumn{2}{c}{\textbf{GR $\uparrow$}} & \multicolumn{2}{c|}{\textbf{MMAU $\uparrow$}} & \textbf{IFrate} \\
    \midrule
    Speech models & \multicolumn{2}{c|}{2.49} & \multicolumn{2}{c|}{85.50} & \multicolumn{2}{c|}{86.00} & \multicolumn{2}{c|}{-} & - \\
    Llama3-8B-Instruct & 7.98 & 5.18 & 86.00 & 83.00 & 86.00 & 85.50 & 48.05 & 49.55 & 92.53 \\
    
    \midrule
    Qwen-Audio-Chat & \textbf{2.49} & 26.25 & 37.00 & 32.50 & 75.00 & 65.50 & 30.03 & 31.53 & 32.98 \\
    Qwen2-Audio-Instruct & 2.73 & \textbf{4.87} & \textbf{67.50} & 39.50 & \textbf{89.00} & 69.50 & 44.44 & 37.54 & 47.11 \\
    
    LTU-AS & 17.31 & 178.69 & 3.00 & 6.50 & 59.50 & 46.00 & 30.33 & 21.02 & 29.19 \\
    SALMONN & 2.64 & 18.48 & 33.00 & 31.50 & 72.50 & 60.00  & 32.13 & 24.02  & 36.89 \\
    \quad \quad \textcolor{gray}{($\alpha=16$)} & \textcolor{gray}{4.87} & \textcolor{gray}{8.84} & \textcolor{gray}{31.50} & \textcolor{gray}{29.00} & \textcolor{gray}{14.50} & \textcolor{gray}{45.50} & \textcolor{gray}{38.74} & \textcolor{gray}{30.63} & \textcolor{gray}{64.26} \\
    \quad \quad \textcolor{gray}{($\alpha=4$)} & \textcolor{gray}{11.07} & \textcolor{gray}{10.43} & \textcolor{gray}{25.00} & \textcolor{gray}{28.00} & \textcolor{gray}{1.50} & \textcolor{gray}{34.50} & \textcolor{gray}{34.83} & \textcolor{gray}{30.33} & \textcolor{gray}{68.06}\\
    \quad \quad \textcolor{gray}{($\alpha=1$)} & \textcolor{gray}{13.42} & \textcolor{gray}{15.82} & \textcolor{gray}{19.50} & \textcolor{gray}{21.00} & \textcolor{gray}{0.00} & \textcolor{gray}{33.00} & \textcolor{gray}{30.33} & \textcolor{gray}{28.83} & \textcolor{gray}{64.86} \\
    BLSP-emo & 56.14 & 25.93 & 35.50 & 38.00 & 0.00 & 28.50 & 27.33 & 29.13 & 60.20 \\
    DiVA & 44.80 & 50.31 & 44.50 & 47.00 & 46.00 & 52.00 & 37.84 & 43.08 & 76.13 \\
    DeSTA2 & 4.87 & 8.92 & 64.50 & \textbf{58.50} & 85.50 & \textbf{85.00} & \textbf{45.35} & \textbf{46.55} & \textbf{89.23} \\
    
    \bottomrule
    \end{tabular}
    \label{tab:speech_results}
\end{table*}

\subsection{Evaluation setup}

For all SLMs and reference systems, we employ greedy decoding with a batch size of 1. In reference systems, we concatenate the structured textual representation with the instruction directly within a conversation. To evaluate model responses, we use a rule-based processor \cite{zhou2023instruction} to automatically verify compliance with constraints in the closed-ended question and creative writing categories. For the chain-of-thought category, we leverage GPT-4o\footnote{gpt-4o-2024-11-20} to assess whether responses demonstrate a coherent and logically structured progression of reasoning.

Beyond our proposed Speech-IFEval, we assess task-level performance, which combines speech perception and instruction-following ability as  conventional evaluation, under both the original instruction and the constraint condition for closed-ended questions. Accuracy across the SER, GR, and MMAU tasks is evaluated using GPT-4o. For ASR evaluation, GPT-4o extracts the most relevant substring from the model’s response and computes the word error rate against the ground-truth transcription. This post-processing step eliminates extraneous context, such as greetings or explanations, ensuring a more precise assessment.

\section{Results}
\subsection{Results on Speech-IFEval}

Table \ref{tab:main_results} presents a comprehensive evaluation of Speech-IFEval, focusing exclusively on instruction-following ability while disregarding speech perception. At first glance, text-only reference systems exhibit outstanding performance, with incremental improvements as models become more advanced. However, despite all SLMs being built upon instruction-following LLMs, most struggle to achieve a reasonable IFrate and exhibit a significant forgetting rate ($\Delta$). This suggests that these models experience significant degradation in instruction-following capabilities.

Early SLMs, such as LTU-AS and SALMONN, represent pioneering efforts in this domain. These models are trained on task-specific speech-instruction pairs to enhance cross-modal integration. However, the construction of these training pairs introduces the risk of overfitting to rigid output structures. While both LTU-AS and SALMONN utilize LoRA adapters to mitigate catastrophic forgetting, their forgetting rates remain notably high (-54.90 and -50.20, respectively).
Qwen-Audio-Chat and its advanced version, Qwen2-Audio-Instruct, are fine-tuned from Qwen-7B, a pre-trained LLM without explicit instruction tuning. Instead, they acquire instruction-following skills during speech-text training. Despite strong performance on speech-instruction benchmarks \cite{sakshi2025mmau, yang-etal-2024-air, 10448257, huang2025dynamicsuperb}, our evaluation reveals notable weaknesses. Qwen2-Audio-Instruct tends to generate direct responses, failing in chain-of-thought tasks, and struggles with formatting conventions, such as double quotation marks and JSON structure. This highlights a limitation of model design without leveraging an instruction-following LLM.
Recently, models like BLSP-emo, DiVA, and DeSTA2 have adopted a similar training approach, using responses from their original LLM counterparts for alignment. Notably, DeSTA2 introduces a speech-text alignment method designed to mitigate the forgetting problem, achieving an impressive IFrate of 89.23. This marks a significant improvement over earlier SLMs and closely matches its text-based LLM counterpart, with a minimal forgetting rate of -3.57.



\subsection{Analysis on task-level performance}

As shown in Table \ref{tab:speech_results}, we evaluate task-level performance of closed-ended questions under two conditions: original instructions (left) and instructions with additional constraints (right). First, SLMs such as Qwen-Audio and SALMONN perform well on specific speech tasks, such as ASR and GR, but exhibit a low instruction-following rate, suggesting that their training is biased toward conventional evaluation metrics. Second, interestingly, comparing the performance under the two conditions, despite the constraints being unrelated to the speech task, most models exhibit significant fluctuations. For example, Qwen2-Audio-Instruct's performance declines from 67.50 to 39.50 for SER and from 89.00 to 69.50 for GR. A closer analysis reveals that most models modify their answer, while others generate hallucinated outputs, such as LTU-AS in ASR task. This indicates that these models are highly sensitive to variations in instructions, underscoring these models' lack of robustness. 
In contrast, DeSTA2, which achieves the highest instruction-following rate (IFrate), exhibits superior consistency and robustness across both  conditions, mirroring the behavior of an LLM cascade pipeline. 


\subsection{Analysis on testing-time LoRA scaling technique}
We apply test-time LoRA scaling \cite{tang2024salmonn} to SALMONN ($\alpha=32$) to reduce the impact of LoRA in LLM during inference. As shown in Tables \ref{tab:main_results} and \ref{tab:speech_results}, this technique significantly improves IFrate, reducing the forgetting rate from -50.20 to -8.12 when $\alpha=4$, but at the cost of task-level performance. This suggests that while test-time pruning can recover some instruction-following ability, maintaining high task-level performance in speech tasks remains challenging. Notably, this finding highlights the limitations of conventional benchmarks that focus solely on task-level performance. Speech-IFEval provides a more systematic analysis of these trade-offs, offering valuable insights for future research.



\section{Conclusion}
We present Speech-IFEval, a comprehensive evaluation framework designed to assess the core capabilities of SLMs by disentangling instruction-following proficiency from speech perception. This separation enables a more precise evaluation of whether and how SLMs retain their textual competencies after speech-text training.
Our findings show that most SLMs fail to preserve instruction-following capabilities, struggle with basic directives, and are highly sensitive to prompt variations. These challenges underscore the inability of current SLM approaches to effectively balance speech perception with robust instruction-following, thereby limiting their adaptability in real-world applications.
Speech-IFEval offers a valuable new standard for evaluating SLMs and offers critical insights to guide future model development.




\bibliographystyle{IEEEtran}
\bibliography{mybib}

\end{document}